\documentclass[journal]{IEEEtran}
\usepackage{graphicx}
\usepackage{cite}

\begin{document}

\title{Interaction between negative and positive index medium waveguides}
\author{Wei Yan,
        Linfang Shen,
        Yu Yuan, and Tzong Jer Yang
\thanks{This work was supported in part by Natural Science Foundation of China
under projects (60277018) and (60531020), and by the financial
support of National Science Council of ROC (NSC 95-2119-M-009-029).}
\thanks{W. Yan, L.F. Shen, and Y. Yuan are with the Department of Information and
Electronic Engineering and Electromagnetics Academy, Zhejiang
University, Hangzhou 310027,P. R. China, T.J. Yang is with the
Department of Electrophysics, National Chiao Tung University,
Hsinchu 30050, Taiwan, ROC (e-mail: lfshen@zju.edu.cn).}}
\maketitle

\begin{abstract}
The coupling between negative and positive index medium waveguides
is investigated theoretically in this paper. A coupled mode theory
is developed for such a waveguide system and its validity is
verified. Interesting phenomena in the coupled waveguides are
demonstrated, which occur in the case when the negative index
medium waveguide in isolation guides its mode backward. A new type
of coupled mode solution that varies exponentially with the
coupling length is found in the special case when the propagation
constants of two individual waveguides are nearly the same. A
coupler operating in this case is insensitive to the coupling
length, and its coupling efficiency can reach $100\%$ as long as
the coupling length is long enough. However, when the propagation
constants of the two individual waveguides differ greatly, the
coupled mode solution is still a periodic function of the coupling
length, but the coupled power is output backward. In addition, the
modes in the composite waveguide system are also studied using the
coupled mode theory, and their fundamental properties are
revealed.
\end{abstract}


\section{Introduction}

Negative index media (NIM) that have simultaneously negative
permittivity and permeability, have attracted intensive interest
recently. These media exhibit several extraordinary effects such
as negative refraction, backward waves, and  evanescent wave
amplification \cite{Veselago, Smith, Pendry}, seeming to challenge
several concepts well established for familiar positve index media
(PIM) in electromagnetism and optics. NIM materials are not
generally found in nature, and thus need to be artificially
constructed. Though the concept of NIM was already proposed by
Vesalogo in 1968 \cite{Veselago}, it was not until 2001 that the
first experimental demonstration of negative index behavior was
accomplished by Shelby et al., with a material made by a
two-dimensional array of repeated unit cells of copper strips and
split-ring resonators \cite{Shelby}. So far, a variety of NIM
structures have been proposed \cite{Shelby, Starr, Zhou, Huangfu,
Chen} in the microwave regime. More recently, NIM structures
operating in the infrared and visible frequency range have also
been reported \cite{Dolling1, Shalaev, Zhang, Dolling2,Dolling3},
e.g., a NIM formed by an array of pairs of parallel gold rods was
demonstrated to have a negative refraction index at a wavelength
of $1.5$ $\mu$m \cite{Shalaev}.

While the study of various NIM structures has been a subject of
growing interest during the past few years, several research works
have involved waveguides with NIM components, and interesting
properties of guidance were reported \cite{Shadrivov, Wu, Peacock}.
For a NIM slab waveguide, the portion of a guided mode inside the
slab has a Poynting vector contradirectional with the direction of
the phase velocity of the mode, while the portion of the guided mode
outside the slab has a Poynting vector parallel to the phase
velocity. The total energy flow of the guided mode may be
codirectional or contradirectional with the phase flow. No
fundamental mode exists in a NIM slab, but it may still support a
single-mode propagation under certain conditions. Moreover, the
interaction between PIM and NIM waveguides was investigated in [17],
and the phenomenon of anti-directional coupling was demonstrated,
which offers a new possibility in the design of future devices and
components. The bound modes of a NIM and PIM composite waveguide
were also studied in [18], [19].

The coupling of waves in planar dielectric waveguides provides a
means of reflectionless signal transfer from one waveguide to
another, thus plays an important role in integrated optics.
Compared to the conventional waveguide system \cite{Hardy1, Haus1,
Snyder}, the coupled NIM and PIM waveguide system receives less
attention. In this paper we will study carefully the coupling
between planar NIM and PIM waveguides. For this purpose, a coupled
mode theory is developed for such a waveguide system. Our analysis
will show that there exist various possibilities of coupled mode
solution for the NIM and PIM waveguide system, and the interaction
of the guided modes of individual waveguides may even be
equivalent to the interference of evanescent modes in the
composite waveguide for particular cases. In Section II, the
coupled mode theory for parallel NIM and PIM waveguides is
formulated, and various coupled mode solutions are discussed in
Section III. In Section IV, the guidance properties of a NIM and
PIM composite waveguide are analyzed. Also, the validity of the
coupled mode theory is examined in this section. Section V
concludes the paper.

\section{The Coupled Mode Equations}

Consider two parallel planar waveguides as illustrated in Fig. 1.
The upper waveguide layer of width $a$ is a PIM with relative
permittivity $\varepsilon_a>0$ and permeability $\mu_a>0$, and the
lower one of width $b$ is a NIM with relative permittivity
$\varepsilon_b<0$ and permeability $\mu_b<0$. The core layers are
separated by a distance of $d$ and they are surrounded by free
space. We assume that the individual PIM waveguide supports only
one guided mode with electric and magnetic fields $\{{\bf e}_1(x)
\exp(j\beta_1z) ,{\bf h}_1(x) \exp(j\beta_1z)\}$ ($\beta_1$ is the
propagation constant of the guided mode), whereas the individual
NIM waveguide may allow multi-mode propagation. The modal fields
of the individual NIM waveguide are denoted by $\{{\bf e}_m(x)
\exp(j\beta_mz) ,{\bf h}_m(x) \exp(j\beta_mz)\}$ ($\beta_m$ are
the propagation constants of the modes), where $m=2, 3, \cdot
\cdot \cdot, N$. We consider an electromagnetic field $\{{\bf E}
(x,z),{\bf H} (x,z)\}$ which satisfies Maxwell's equations plus
the boundary conditions of the entire structure. Its transverse
field can be written as
\begin{eqnarray}
&&{\bf E}_t (x,z) = \sum\limits_{m = 1}^N {u_m (}z)\left[{\bf
e}_{mt}(x)+\delta {\bf e}_m(x)\right] ,\\
&&{\bf H}_t (x,z) = \sum\limits_{m = 1}^N {u_m (}z)\left[{\bf
h}_{mt}(x)+\delta {\bf h}_m(x)\right],
\end{eqnarray}
where $\delta {\bf e}_m$ and $\delta {\bf h}_m$ ($m=1$, $2$,
$\cdot \cdot \cdot$, $N$) are the corrections of the modal field
in the other guide due to the induced polarization. The correction
fields $\delta {\bf e}_m$ are introduced as in Ref. [22], and
$\delta {\bf h}_m$ are treated in a similar manner. The
longitudinal components of the vectorial field are expressed in
the form
\begin{eqnarray}
&&{\bf E}_z (x,z) = \sum\limits_{m = 1}^N {u_m
(}z)\frac{{\varepsilon_m(x) }} {\varepsilon(x)
}{\bf e}_{mz}(x),\\
&&{\bf H}_z (x,z) = \sum\limits_{m = 1}^N {u_m (}z)\frac{{\mu_m(x)
}} {\mu(x)}{\bf h}_{mz}(x),,
\end{eqnarray}
where $\varepsilon_1=\varepsilon^{(a)}$, $\mu_1=\mu^{(a)}$;
$\varepsilon_m=\varepsilon^{(b)}$, $\mu_m=\mu^{(b)}$ for $m\ge2$.
$\{\varepsilon,\mu\}$, $\{\varepsilon^{(a)},\mu^{(a)}\}$, and
$\{\varepsilon^{(b)},\mu^{(b)}\}$ represent the profiles of the
entire structure, and of the individual PIM and NIM waveguides,
respectively.

\begin{figure}
\centering
\includegraphics[width=8cm]{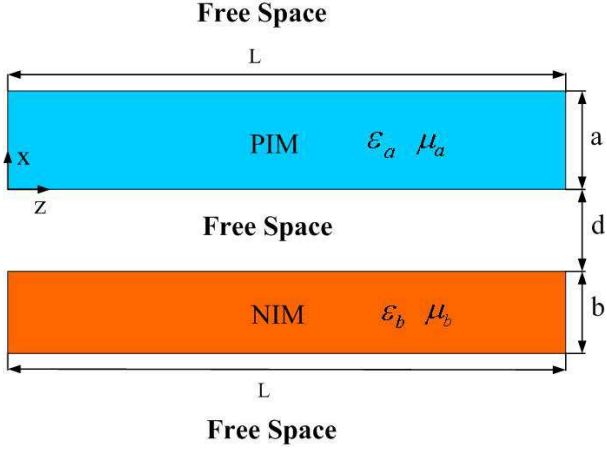}
\caption{Schematic of the coupled waveguide structure.}
\label{fig_1}
\end{figure}

We utilize a variational principle for the coupled system under
consideration following Haus et al. [21], and obtain the
differential equations in the form
\begin{eqnarray}
P\frac{d}{{dz}}U =  - jPBU - jKU,
\end{eqnarray}
where $U = \mathrm{col}[u_1 ,u_2, \cdot \cdot \cdot, u_N]$, and $B
= \mathrm{diag}[\beta _1 ,\beta _2, \cdot \cdot \cdot, \beta_N]$.
The matrices $P$ and $K$ have elements
\begin{eqnarray}
P_{mn}&=& {\bar P}_{mn}+\left(\delta P_{mn}+\delta
P_{nm}^*\right)+\Delta P_{mn},\quad\\
K_{mn}&=&{\bar K}_{mn} + \delta K_{mn}+ (\beta _m  - \beta
_n)\delta P_{mn}+\Delta K_{mn},\quad
\end{eqnarray}
with
\begin{eqnarray}
{\bar P}_{mn}&=& \int{[{\bf e}_{mt} ^ *\times {\bf h}_{nt}+{\bf
e}_{nt} \times
{\bf h}_{mt}^*] \cdot{\kern 1pt} {\kern 1pt} \widehat z dx},\\
\bar K_{mn} &=& \omega \int [\Delta \varepsilon _n {\bf e}_m ^*
\cdot {\kern 1pt} {\kern 1pt}  {\bf e}_n  - (\Delta \varepsilon_m
\Delta \varepsilon _n / \varepsilon)
{\bf e}_{zn}\cdot {\kern 1pt} {\kern 1pt} {\bf e}_{zm}^* \nonumber\\
&+& \Delta \mu _n{\bf h}_m^* \cdot {\kern 1pt} {\kern 1pt} {\bf
h}_n-(\Delta \mu _m \Delta \mu _n/\mu){\bf h}_{zn} \cdot {\kern
1pt} {\kern 1pt}
 {\bf h}_{zm}^*]dx, \nonumber\\
\end{eqnarray}
and
\begin{eqnarray}
\delta P_{mn}  &=& \int{[\delta {\bf e}_n  \times {\bf h}_{mT} ^ *
+ {\bf e}_{mT} ^* \times \delta {\bf h}_n ] \cdot {\kern 1pt}
{\kern 1pt} \widehat z
{\kern1pt}dx},\\
\Delta P_{mn}  &=& \int{[\delta {\bf e}_m^ * \times\delta {\bf
h}_n+\delta {\bf e}_n
\times \delta {\bf h}_m^*] \cdot{\kern 1pt} {\kern 1pt} \widehat zdx},\\
\delta K_{mn}  &=& \int[\Delta \varepsilon _m {\bf e}_m ^* \cdot
{\kern 1pt} {\kern 1pt} \delta {\bf e}_n + \Delta \varepsilon _n
\delta {\bf e}_m ^
* \cdot {\kern 1pt} {\kern 1pt} {\bf e}_n + \Delta \mu _n \delta {\bf h}_m ^ *
 \cdot {\kern 1pt} {\kern 1pt}
{\bf h}_n \nonumber\\
 &+& \Delta \mu _m {\bf h}_m ^ *   \cdot {\kern 1pt} {\kern
1pt} \delta
{\bf h}_n] dx,\\
\Delta K_{mn}&=&\int[\varepsilon \delta {\bf e}_m ^ *   \cdot
{\kern 1pt} {\kern 1pt} \delta {\bf e}_n+ \mu \delta {\bf h}_m ^ *
\cdot {\kern 1pt} \delta {\kern 1pt} {\bf h}_n+(\nabla _t \times
\delta
{\bf h}_n ) \cdot {\kern 1pt} {\kern 1pt} \delta {\bf e}_m ^ *\nonumber\\
&-& (\nabla _t \times \delta {\bf e}_n ) \cdot {\kern 1pt} {\kern
1pt} \delta {\bf h}_m ^* ] dx,
\end{eqnarray}
where $\Delta \varepsilon _m=\varepsilon-\varepsilon_m$ and
$\Delta \mu _m=\mu-\mu_m$. Note that $\Delta P_{mn}=\delta
K_{mn}=\Delta K_{mn}=0$ for $m\ne n$.

The above coupled-mode equations are similar to those derived by
Haus et al. in [23], but our formulation considers a more general
case, in which the permeability of the coupled waveguides is
allowed to vary with the transverse coordinate ($x$). Furthermore,
our formulation includes the factor $\varepsilon_i/\varepsilon$
$\left(\mu_i/\mu\right)$ in the expression for $E_z$ ($H_z$). We
let $H=PB+K=\bar H+\delta H$, here $\bar H=\bar P \bar B+\bar K$,
then Eq. (5) can be written as $PdU/dz=-jHU$. In the case if the
waveguide system is lossless, $\bar H$ can be proved to be
Hermitian in the same manner as in [10]. As the matrix $\delta H$
has off-diagonal elements $\delta H_{mn}=\beta_m \delta
P_{mn}+\beta_n \delta P_{nm}^*$, thus the matrix $H$ is also
Hermitian for the lossless case. So the power conservation holds
for our formulation \cite{Hardy2}.

\section{Coupled Mode Solutions}

Let us consider the wave propagation and coupling along two lossless
planar waveguides over a finite length $L$ (see Fig. 1). For
simplicity, we assume that the individual NIM waveguide also
supports a singled mode (i.e., $N=2$). The single-mode operation for
a certain polarization in the NIM waveguide can be achieved through
suppressing the appearance of surface modes. We consider an initial
power to be injected into the PIM waveguide at $z=0$. The excited
mode in the NIM waveguide will have the same direction of phase flow
as the guided mode in the PIM waveguide, i.e., $\beta_2>0$. As the
energy flows of the mode in the core and cladding of the NIM
waveguide are in the opposite directions, the total energy flow may
be codirectional or contradirectional with the phase velocity of the
mode, i.e., $\bar P_{22}>0$ or $\bar P_{22}<0$. The coupled power in
the NIM waveguide will be output at the waveguide end $z=L$ if $\bar
P_{22}>0$. Otherwise the coupled power will be output at the other
end $z=0$, if $\bar P_{22}<0$. Correspondingly, the boundary
condition is $u_2=0$ at $z=0$ for $\bar P_{22}>0$, and it becomes
$u_2=0$ at $z=L$ for $\bar P_{22}<0$.

We rewrite Eq. (5) in the form
\begin{eqnarray}
\frac{d}{{dz}}U =  - jBU - jCU,
\end{eqnarray}
where $C=P^{-1}K=P^{-1}H-B$, with the superscript "$-1$" denoting
an inverse matrix. The coefficients $C_{mm}$ ($m=1$, $2$) in Eq.
(14) are equivalent to a modification of the parameters $\beta_m$,
and our analysis shows that $|C_{mm}| \ll\beta_m$ in general. The
coupling coefficients $C_{12}$ and $C_{21}$ are especially
interesting. These coupling coefficients are given by $ C_{12} =
(P_{22} H_{12} - P_{12} H_{22} )/(P_{11} P_{22}  - P_{12} P_{21})$
and $ C_{21} = (P_{11} H_{21}  - P_{21} H_{11} )/(P_{11} P_{22}  -
P_{12} P_{22} )$, where $H_{11}=P_{11}\beta_1+K_{11}$ and
$H_{22}=P_{22}\beta_2+K_{22}$. Since $|P_{12} P_{21}|\ll |P_{11}
P_{22}|$, $|K_{22}|\ll |P_{22}\beta_2|$, and $P_{11}\approx \bar
P_{11}$ we find $C_{12}\approx (H_{12}-P_{12}\beta_2)/{\bar
P}_{11}$. Similarly, we obtain $C_{21}\approx
(H_{21}-P_{21}\beta_1)/{\bar P}_{22}$. Note that $H_{12}=H_{21}$
and $P_{12}=P_{21}$ in the lossless case. If $\beta_1$ and
$\beta_2$ differ negligibly, i.e., $\beta_1 \approx \beta_2$, then
$C_{12}$ and $C_{21}$ have the same sign for $\bar P_{22}>0$, or
they have opposite signs for $\bar P_{22}<0$, which never happens
for coupled conventional waveguides. As an example, the coupling
coefficients as a function of the NIM layer thickness $b$ are
illustrated in Fig. 2(a), and the value of ($\beta_2-\beta_1$) is
shown in Fig. 2(b). The guided wave is assumed to be TE-polarized,
and the parameters of the two-waveguide system are as follows:
$\varepsilon_1=2.1$, $\mu_1=1$, and $a=0.3035\lambda$,
corresponding to the waveguide parameter $V=1$ for the PIM
waveguide; $\varepsilon_2=-1.8$, and $\mu_2=-1$; $d=2a$. As seen
from Fig. 2(a), in the region of $0.36\lambda<b<0.55\lambda$ where
$\bar P_{22}>0$, the coupling coefficients are both positive; but
they have opposite signs in the region of
$1.12\lambda<b<1.63\lambda$ where $\bar P_{22}<0$. The shaded
areas in Fig. 2 indicate the regions where the single-mode
propagation in the individual NIM waveguide is not available or
only the surface mode exists. In what follows, we concentrate on
the analysis of the coupling between PIM and NIM waveguides for TE
polarization (results for TM polarization can be obtained from
duality). We will consider two cases for waveguide system: (i)
${\bar P}_{22}<0$; (ii) ${\bar P}_{22}>0$.

\begin{figure}[h]
\includegraphics[width=8cm]{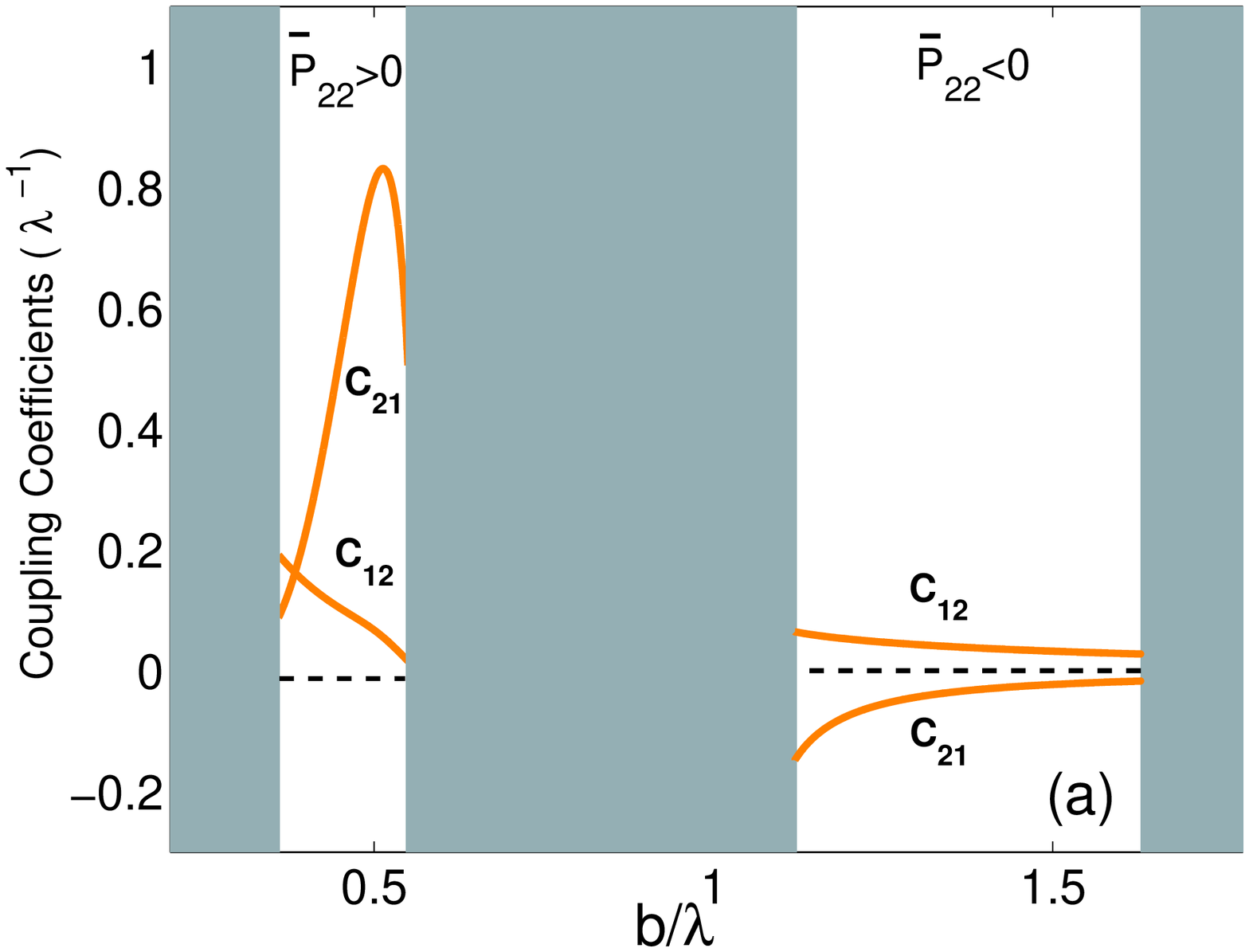}
\includegraphics[width=8cm]{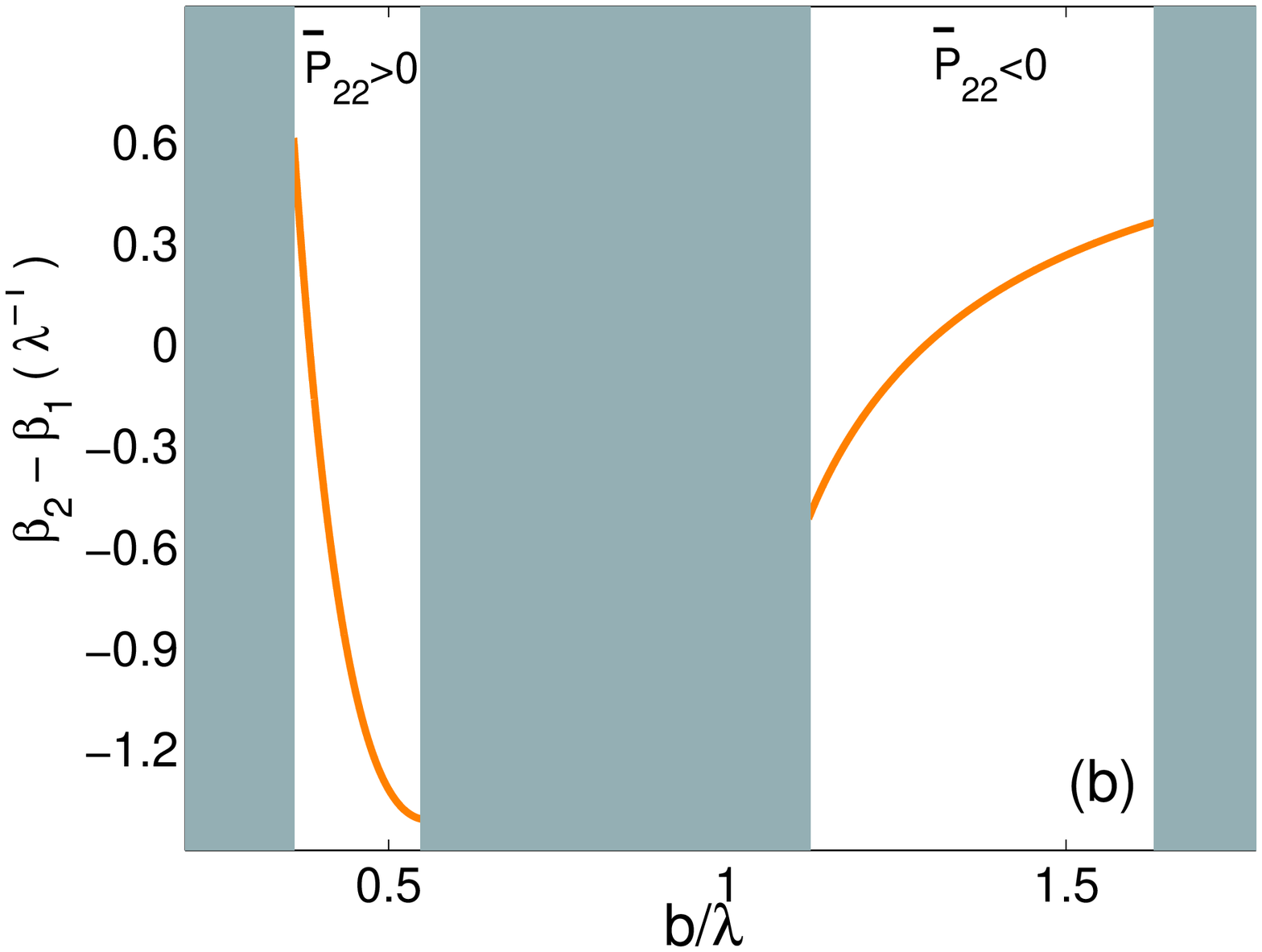}
\caption{(a) Coupling coefficients versus the NIM layer thickness
$b$. (b) Propagation constant difference $(\beta_2-\beta_1$)
versus the NIM layer thickness $b$. The shaded areas indicate the
regions where the single-mode propagation in the individual NIM
waveguide is not available.}
\end{figure}

 A. Case (i): ${\bar P}_{22}<0$

The solution satisfying $u_1=u_1(0)$ at $z=0$ and $u_2=0$ at $z=L$
is shown to be
\begin{eqnarray}
u_1 (z) &=& A\{\psi  -j \Delta \tanh [\psi (z - L)]\}e^{j\phi z},\\
u_2 (z) &=& jAC_{21}\tanh [\psi (z - L)]e^{j\phi z}
\end{eqnarray}
for the case if $\Delta ^2 \le - C_{12}C_{21}$, or else it becomes
\begin{eqnarray}
u_1 (z) &=&A\{\psi \cos [\psi (z - L)] - j\Delta \sin [\psi (z -
L)]\}e^{j\phi z},\nonumber\\
\\
u_2 (z) &=& A C_{21}\sin [\psi (z - L)]e^{j\phi z},
\end{eqnarray}
where $\psi = \sqrt {| \Delta ^2 + C_{12}C_{21}|}$, $\Delta =
(\beta _1 ^{'} - \beta _2^{'})/2$, and $\phi  = (\beta _1 ^{'}  +
\beta _2 ^{'} )/2$ with $\beta_m^{'}=\beta_m+C_{mm}$ ($m=1$, $2$).
Since $C_{12}C_{21}<0$ at least when $\beta_1 \approx \beta_2$
(i.e., $\Delta \approx 0$), the first type of solution, i.e., Eqs.
(15) and (16), is possible to occur for the case with ${\bar
P}_{22}<0$. Note that Eqs. (15) and (16) are a new type of the
solution for coupled mode equations, which never happens for a
conventional waveguide system.

The coupling efficiency (or the fraction of the coupled power),
which describes the net power transfer between the two waveguides,
is determined by $\eta=|\bar P_{22}|u_2(0)^2/\bar P_{11}u_1(0)^2$.
In the case of $\Delta ^2 \le - C_{12}C_{21}$, we have $\eta=|\bar
P_{22}|C_{21}^2/\bar P_{11}[\psi ^2 \coth^2(\psi L) + \Delta ^2]$,
thus the output power by the NIM waveguide increases as the length
$L$ increases, and it approaches its maximum of $\eta_{max}=|\bar
P_{22}C_{21}/\bar P_{11}C_{12}|$ as $L> 1/\sqrt {|C_{12}C_{21}|}$.
Evidently, $\eta_{max}$ can reach nearly $100\%$ when $\beta_1
\approx \beta_2$. In the other case if $\Delta ^2 \ge -
C_{12}C_{21}$, where $\beta_1$ and $\beta_2$ differ greatly, the
coupling efficiency becomes $\eta=|\bar P_{22}|C_{21}^2/\bar
P_{11}[\psi ^2 \cot^2(\psi L) + \Delta ^2]$. In this case we find
that $\eta$ varies periodically with $L$, and it has a maximum of
$\eta_{max}=|\bar P_{22}C_{21}^2/\bar P_{11}\Delta^2|$ at the
lengths of $L=(n-1/2)\pi/\psi$, here $n$ is an integer. Since
$\Delta ^2 \ge - C_{12}C_{21}$, we have $\eta_{max} \le |\bar
P_{22}C_{21}/\bar P_{11}C_{12}|$. For this case our numerical
analysis indicates that $\eta_{max}$ is always less than $1$. To
illustrate the typical coupling behaviors, Fig. 3 shows the
fraction of the coupled power for TE modes as a function of $L$
for the waveguide systems with $b= 1.2947\lambda$ and
$1.228\lambda$. The other parameters of the waveguide systems are
the same as in Fig. 2. The propagation constant of the individual
NIM waveguide has $\beta_2=\beta_1$ for $b=1.2947\lambda$ and
$\beta_2=0.9816\beta_1$ for $b=1.228\lambda$, corresponding to two
typical cases discussed above. Here, we should point out that a
NIM is inherently lossy, i.e., both $\epsilon_b$ and $\mu_b$ have
an imaginary part. Thus, the propagation constant ($\beta_2$) of a
guided mode in the individual NIM waveguide also has an imaginary
part. In the case if the NIM is severely lossy, the new type of
coupled mode solution will not occur even when $\beta_1$ is equal
to the real part of $\beta_2$. But such a NIM with considerable
loss is not suitable for application in devices. By using low loss
materials to structure NIM, it is possible to reduce the loss of a
NIM to very low level at certain frequencies, which is the case of
our interest.

\begin{figure}[h]
\includegraphics[width=8cm]{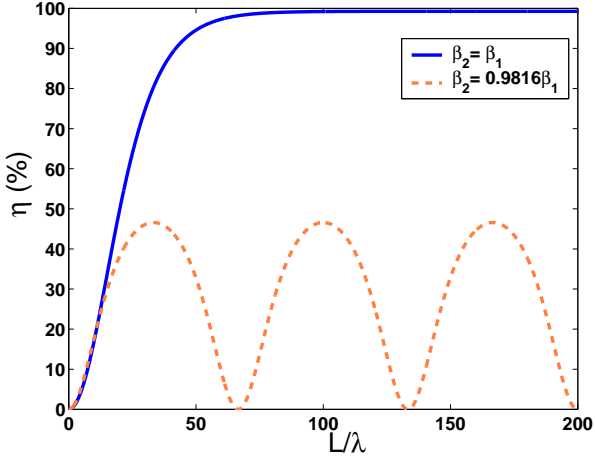}
\caption{Fraction of coupled power $\eta$ ($ \% $) versus the
coupling length $L$.}
\end{figure}

B. Case (ii): ${\bar P}_{22}>0$

The solution with $u_1=u_1(0)$ and $u_2=0$ at $z=0$ is given by
\begin{eqnarray}
 u_1 (z) &=& A[\psi \cos (\psi z) - i\Delta \sin (\psi z)]e^{i\phi
 z},\\
 u_2 (z) &=& AC_{21}\sin (\psi z)e^{i\phi z},
\end{eqnarray}
where $\phi$, $\Delta$, and $\psi$ are defined above. In this case
the coupling efficiency is determined by $\eta=|\bar
P_{22}|u_2(L)^2/\bar P_{11}u_1(0)^2$, and it follows that
$\eta=\bar P_{22}C_{21}^2\sin^2(\psi L)/\bar P_{11}\psi ^2$, which
is identical to that for two parallel conventional waveguides. The
maximum of the coupling efficiency is $\eta_{max}=\bar
P_{22}C_{21}^2/\bar P_{11}\psi ^2$, occurring at the lengths of
$L=\pi (n-1/2)\psi$, where $n$ is an integer, and it reaches
nearly $100\%$ when $\beta_2$ is equal to $\beta_1$.

\section{Guidance of the Composite Waveguide}

It is interesting to analyze the guidance characteristic of the
composite two-core waveguide, as the coupling between guided modes
of two individual waveguides may always be viewed as the beating
of the modes of the composite structure. The propagation constants
of the modes in the composite waveguide are easily derived from
Eq. (14), and are given by
\begin{eqnarray}
\beta_{\pm} = \frac{{\beta _1^{'}  + \beta _2 ^{'} }}{2} \pm
\frac{{\sqrt {(\beta _1 ^{'}  - \beta _2 ^{'} )^{2}  + 4C_{12}C_{21}
} }}{2},
\end{eqnarray}
where $\beta_m^{'}$ ($m=1$, $2$) are defined in the previous
section. It is known that the exact dispersion relation for the
planar composite waveguide can be derived analytically for both TE
and TM polarizations. The electric field ($\bf E$) for TE modes in
the five regions of the waveguide structure (see Fig. 1) can be
written as
\begin{eqnarray}
&&{\bf E}_1  = \hat y A_1 \exp ( - \alpha x + i\beta z),\quad\quad
x >
a \,(\rm {free\,space})\nonumber\\
&&{\bf E}_2  = \hat y\left[A_2 \exp (ik_P x) + B_2 \exp ( - ik_P
x)\right]\exp(i\beta z),\nonumber\\
&&\quad\quad\quad\quad\quad\quad\quad\quad\quad\quad\quad\quad\quad\quad 0 < x \le a \,(\rm{PIM})\nonumber\\
&&{\bf E}_3  = \hat y\left[A_3 \exp ( - \alpha x) + B_3 \exp
(\alpha x)\right]\exp(i\beta z),\nonumber\\
&&\quad\quad\quad\quad\quad\quad\quad\quad\quad\quad\quad\quad - d < x \le 0 \,(\rm{free\,space})\nonumber\\
&&{\bf E}_4  = \hat y\left[A_4 \exp (ik_N x) + B_4 \exp ( - ik_N
x)\right]\exp(i\beta z),\nonumber\\
&&\quad\quad\quad\quad\quad\quad\quad\quad\quad\quad\quad- (d + b) < x \le  - d \,(\rm{NIM})\nonumber\\
&&{\bf E}_5  = \hat y A_5 \exp (\alpha x + i\beta z),\quad x \le -
(d+ b)\,(\rm{free\,space})\nonumber
\end{eqnarray}
where $ \alpha = \sqrt {\beta ^2  - k_0^2 }$, $ k_P = \sqrt
{\varepsilon _a \mu _ak_0^2 - \beta ^2 }$, $ k_N = \sqrt
{\varepsilon _b \mu _bk_0^2 - \beta ^2 }$, and $k_0$ is the wave
number in free space. The nonzero components of the magnetic field
($\bf H$) can be obtained straightforwardly from $\bf E$. The
dispersion relation of TE modes is determined by imposing matching
conditions on the parallel components of $\bf E$ and $\bf H$ at
the interfaces $x=-(b+d)$, $-d$, $0$, and $a$. Following this
procedure yields
\begin{eqnarray}
[(f_1  - 1/f_1 ) - 2\cot (k_P a)][(f_2  - 1/f_2 ) - 2\cot (k_N b)]e^{\alpha d}  \nonumber \\
= (f_1  + 1/f_1 )(f_2  + 1/f_2 )e^{ - \alpha
d}\quad\quad\quad\quad\quad
\end{eqnarray}
with $ f_1  = k_P /(\alpha \mu _a )$ and $ f_2  = k_N /(\alpha \mu
_b )$. This equation is the exact dispersion relation for TE modes
in the composite waveguide. The exact dispersion relation for TM
modes can be obtained by substitution of $\mu \leftrightarrow
\varepsilon $ in Eq. (22). However, it is very difficult to solve
analytically the transcendental equation (22). In contrast, the
formula (21) enables one to gain more insight into the
characteristic of the waveguide system. In the following we focus
on the analysis of TE modes in the composite waveguide (results
for TM modes can be obtained from duality). We will show that Eq.
(21) is a good approximation to Eq. (22). Without losing
generality, the interacted modes in the individual PIM and NIM
waveguides are assumed to have positive propagation constants,
thus the real part of the propagation constant of each related
mode in the composite waveguide is also positive. All results
calculated from Eq. (21) will be compared with the exact results
obtained from Eq. (22). This provides an approach to examine the
validity of the coupled mode theory.

We first consider the waveguide system with $\bar P_{22}<0$. In
the special case with $\beta_1\approx\beta_2$, where
$C_{12}C_{21}<0$, one sees from Eq. (21) that the propagation
constants have an imaginary part and be conjugate for any
separation distances $d$. Note that both the NIM and PIM in the
waveguide system are assumed to be lossless here. So these modes
in the composite waveguide are evanescent waves. The appearance of
the evanescent modes is only a consequence of the special coupling
between the PIM and NIM waveguides, which corresponds to the new
type of coupled mode solution. In the case if $\beta_1$ and
$\beta_2$ differ substantially, the propagation constants may also
have an imaginary part for small $d$, namely, when the coupling
between the individual waveguides is strong. However, as $d$ is so
increased that $|C_{12}C_{21}|\le (\beta_1-\beta_2)^2$ , the
propagation constants will become real. In this case both modes in
the composite waveguide are propagating waves. In the case if
$\beta_1$ and $\beta_2$ differ greatly, the propagation constants
are always real for any $d$. To illustrate these behaviors, Fig. 4
shows the propagation constants for the composite waveguides with
different NIM layer thicknesses $b=1.2947\lambda$, $1.228\lambda$
and $1.1559\lambda$. The other parameters are the same as in Fig.
2. The propagation constants of the individual NIM waveguide are
$\beta_2=\beta_1$, $0.9816\beta_1$, and $0.9531\beta_1$ for the
three cases, respectively. In Fig. 4, the exact results (labelled
by "Exact") for the three cases are also included for comparison
with the results (labelled by "CMT") obtained from Eq. (21), and
the excellent agreement is observed.

\begin{figure}[htbp]
\centering
\includegraphics[width=7cm]{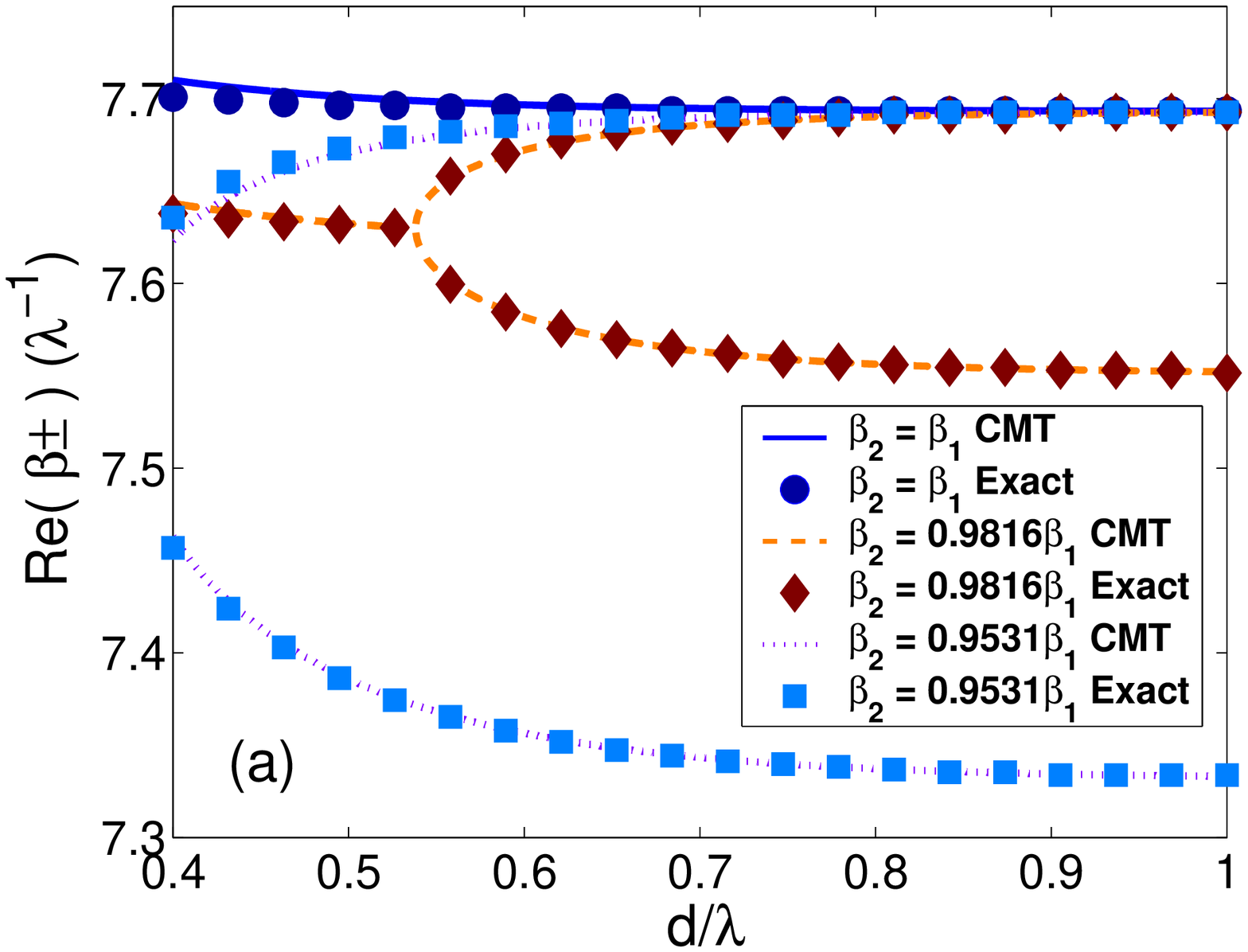}
\includegraphics[width=7cm]{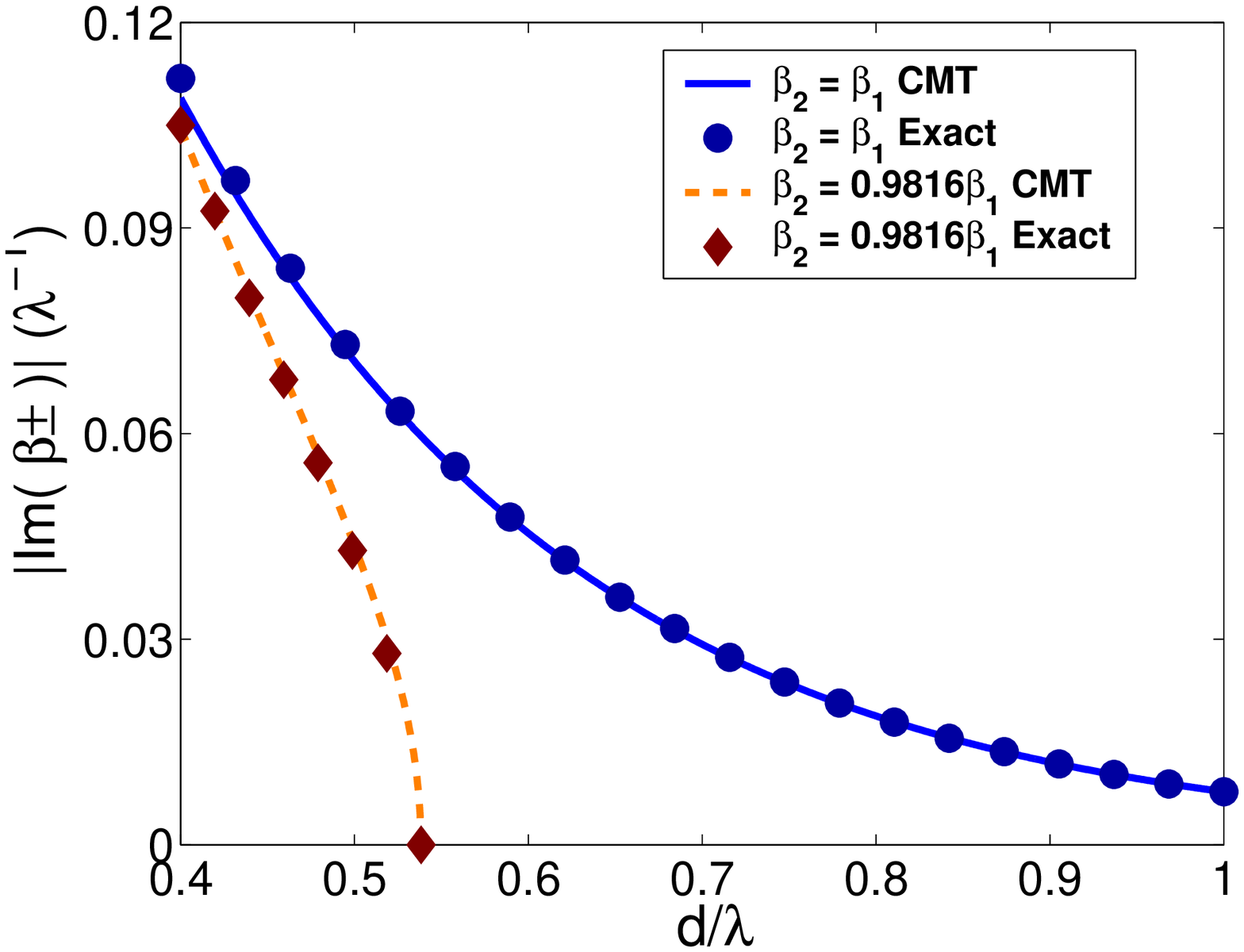}
\caption{ Propagation constants $\beta_{\pm}$ versus the distance
$d$ between the core layers for the composite waveguides with
various NIM layer thicknesses. The NIM waveguide in isolation
guides a single mode backward ($\bar P_{22}<0$) for all cases. (a)
Real parts of $\beta_{\pm}$; (b) imaginary parts of $\beta_{\pm}$.
The lines represent the results obtained from coupled mode theory,
and the marks represent the results calculated from Eq. (22).}
\end{figure}

Next, we consider the waveguide system with $\bar P_{22}>0$. In
this case, our numerical analysis indicates that $C_{12}C_{21}>0$,
and the propagation constants of the modes of the composite
waveguide are always real, as illustrated in Fig. 5, where the
thicknesses of the NIM layers are $b=0.3832\lambda$,
$0.3895\lambda$ and $0.4007\lambda$, corresponding to the
propagation constants of the guided modes in the individual NIM
waveguides $\beta_2=\beta_1$, $0.9816\beta_1$, and
$0.9531\beta_1$, respectively. The other parameters are the same
as in Fig. 2. The exact results of the propagation constants for
the composite waveguides are also included in Fig. 5 for
comparison. One sees that the results obtained from Eq. (21) are
in good agreement with the exact results.

\begin{figure}[htbp]
\centering
\includegraphics[width=7cm]{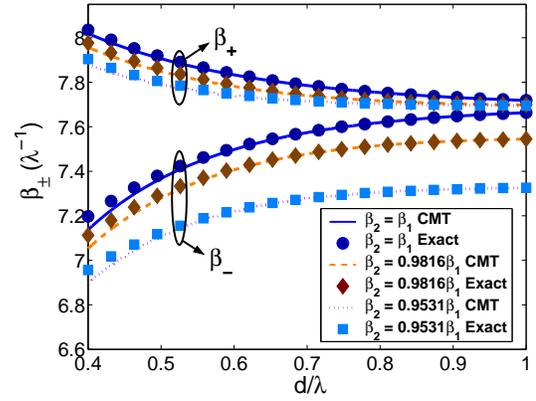}
\caption{Propagation constants $\beta_{\pm}$ versus the distance
$d$ between the core layers for the composite waveguides with
various NIM layer thicknesses. The NIM waveguide in isolation
guides a single mode forward ($\bar P_{22}>0$) for all cases. The
lines represent the results obtained from coupled mode theory, and
the marks represent the results calculated from Eq. (22).}
\end{figure}

Finally we analyze qualitatively the total energy flow of the
modes in the composite waveguide, which is given by $S=U^+ PU$,
where the superscript $"+"$ denotes the conjugated transpose of
matrix. For the waveguide system with $\bar P_{22}>0$, $P$ is a
positive Hermitian matrix, which leads to $S>0$, indicating that
the modes of the composite waveguide are always forward modes,
i.e., the total energy flows of the modes are codirectional with
their phase flows. Therefore, the guided modes shown in Fig. 5 are
all forward modes. This is verified by the exact calculation of
the total energy flows of the modes in the composite waveguide. On
the other hand, for the waveguide system with $\bar P_{22}<0$, $P$
is an indefinite Hermitian matrix, and it can be written as $ P =
R^ + J R $, where $J=\mathrm{diag}[1, -1]$, and $R^ +   = [{\bf
r}_1,{\bf r}_2] {\kern 1pt} \,\mathrm{diag}[|\sqrt {|\lambda _1 |}
,\sqrt {|\lambda _2 |} ] $, with $\lambda _n$ and ${\bf r}_n$
($n=1$, $2$) being the eigenvalues and orthogonal eigenvectors of
$P$, respectively. Let $V=RU=\mathrm{col}[v_1,v_2]$, then the
coupled mode equations are expressed as $dV/dz= jTV$, where $T$ is
an anti-Hermitian matrix, given by $ T = J[R^ + ]^{ - 1} (PB +
K)R^{ - 1}=J[R^ + ]^{ - 1} HR^{ - 1}$, with $T_{12}=-T_{21}^{*}$.
We also have $S=V^+ JV=|v_1|^2-|v_2|^2=(|v_1/v_2|^2-1)|v_2|^2$.
Solving the coupled mode equations, we find
\begin{eqnarray}
\beta _ \pm   = \frac{{(T_{11}  + T_{22} ) \pm \sqrt {(T_{11}  -
T_{22} )^2  - 4|T_{12}|^2 } }}{2},
\end{eqnarray}
and
\begin{eqnarray}
|v_1 /v_2 | = |\xi \pm \sqrt {\xi^2  - 1} | ,
\end{eqnarray}
where $\xi = |\left( {T_{11}-T_{22} } \right)/2T_{21} |\ge 0$.
Note that it is not necessary to take the same sign simultaneously
in Eqs. (23) and (24). As seen from Eq. (23), when $\xi\le1$, the
propagation constants of two modes in the composite waveguide have
an imaginary part. In this case we have $ |v_1 /v_2|=1$ and then
$S=0$, which indicates that the modes in the composite waveguide
are indeed evanescent waves, though they have a nonzero real part
of the propagation constants. In the case when $\xi> 1$,
$\beta_{\pm}$ are both real, i.e., the modes in the composite
waveguide are both propagating waves. In this case one has $|v_1
/v_2 |>1$ and then $S>0$, if we take $"+"$ in Eq. (24), while
$|v_1 /v_2 |<1$ and then $S<0$, if choosing $"-"$ in Eq. (24).
Thus, one of the two modes in the composite waveguide is a forward
mode, and the other must be a backward mode, whose total energy
flow is contradirectional with its phase flow. Since $ |P_{12}
|,\,|P_{21} |\, \ll \,|P_{11} |,\,|P_{22} |$ in general cases
(i.e., except in the case of super strong coupling), the matrix
$P\approx \mathrm{diag}[P_{11},P_{22}]$, thus we have
approximately $ R \approx \mathrm{\mathrm{diag}}[\sqrt {|P_{11} |}
,\sqrt {|P_{22} |} ]$. Substituting this equation into the
expression of $T$, we find that $T_{11}\approx \beta_1 $,
$T_{22}\approx \beta_2$. So we infer that $S>0$ for $\beta_+$ and
$S<0$ for $\beta_-$, if $\beta_1>\beta_2$, otherwise  $S<0$ for
$\beta_+$ and $S>0$ for $\beta_-$, if $\beta_1<\beta_2$.
Therefore, in Fig. 4, the guided modes with propagation constant
$\beta_+$ are forward modes, while the ones with propagation
constant $\beta_-$ are backward modes. This is also verified by
the exact numerical calculation of the total energy flow.

\section{Discussion and Conclusions}
The wave propagation and coupling in parallel planar NIM and PIM
waveguides have been studied theoretically. The coupled mode
equations for such a waveguide system has been developed, which
has been verified through calculation of the propagation constants
of the modes in the composite waveguide. It has been shown that if
the NIM waveguide in isolation guides its mode forward, the
properties of the NIM and PIM waveguide system are similar to
those for a conventional waveguide system. However, in the case if
the NIM waveguide guides its mode backward, interesting phenomena
then appear. When the propagation constants of the individual
waveguides are nearly equal, there exist only evanescent modes in
the composite waveguide. The solution of the coupled mode
equations for this case varies exponentially with the coupling
length, which never happens for a conventional waveguide system. A
coupler operating in this case is very easy to fabricate, as its
coupling efficiency is insensitive to the coupling length. When
the propagation constants $\beta_1$ and $\beta_2$ differs
significantly, the modes in the composite waveguide may change
from evanescent waves to propagating ones as $d$ increases. In the
latter case the solution of the coupled mode equations is also a
periodic function of the coupling length, but the coupled power is
output in a direction opposite to that of input power.
Correspondingly, the modes in the composite waveguide are
propagating waves, and one of the modes is a forward mode while
the other is a backward one. Finally we should indicate that as
the NIM medium is strongly dispersive, the various phenomena
mentioned above may happen in the same waveguide system at
different frequencies.

\begin{biography}[{\includegraphics[width=1in,height=1.25in,clip,keepaspectratio]{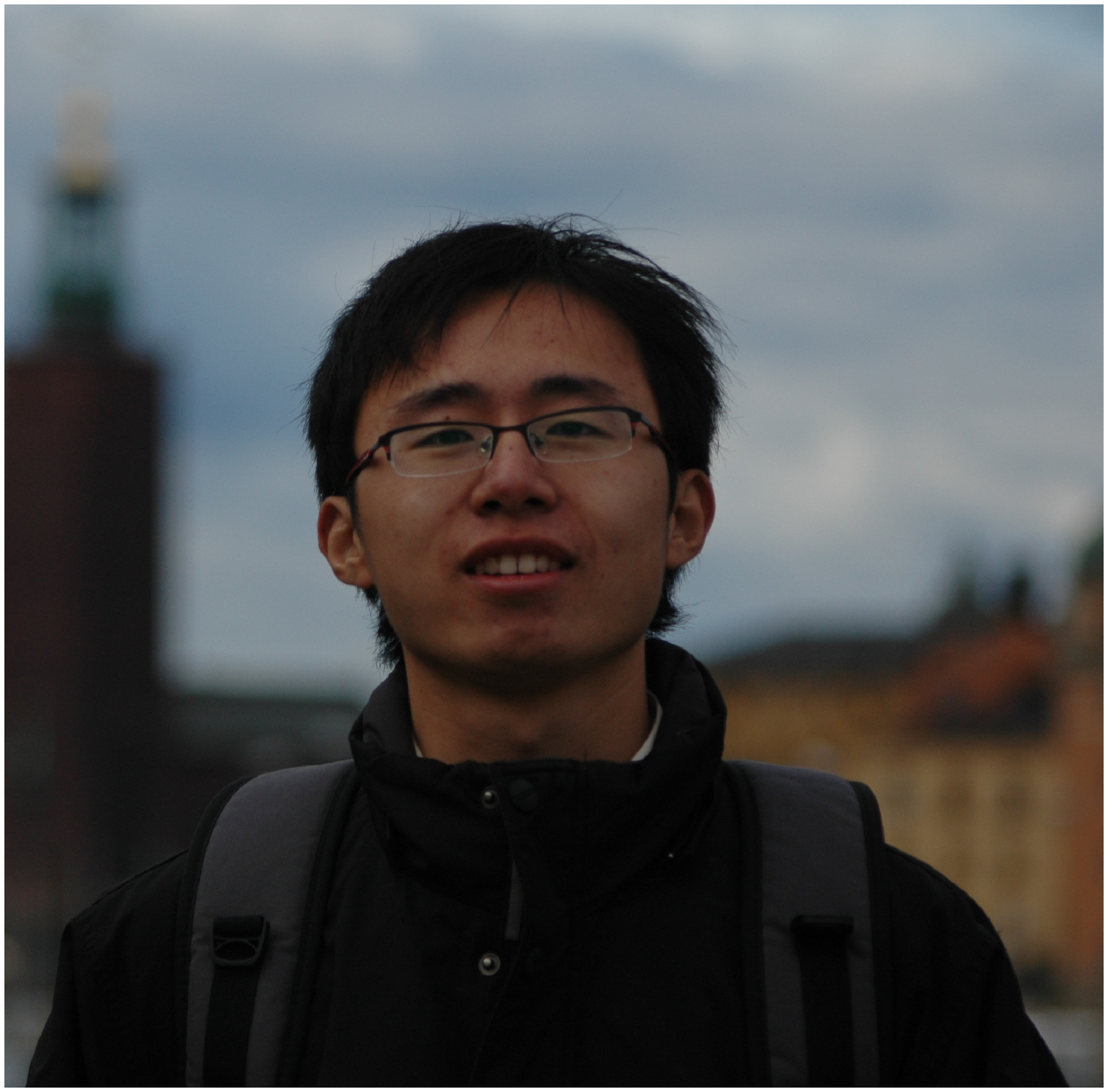}}]{Wei Yan}
was bron in Jiangsu, China, in 1983. He received the B.S. degree in
2004 from Zhejiang University, Hangzhou, China, where he is
currently working toward the M.S. degree. His current research
interests include photonic crystals and metamaterials.
\end{biography}
\begin{IEEEbiographynophoto}{Linfang Shen}
was born in Zhejiang, China, in 1965. He received the B.S. degree in
physics from Peking University, Beijing, China, in 1986, the M.S.
degree in plasma physics from Institute of Plasma Physics, Academy
of Science of China, Hefei, China, in 1989, and the Ph.D. degree in
Electronic Engineering from University of Science and Technology of
China, Hefei, China, in 2000. He is currently an Associate Professor
with Zhejiang University. His present research interests are in
photonic crystals, photonic crystal fibers, and metamaterials.
\end{IEEEbiographynophoto}
\begin{IEEEbiographynophoto}{Yu Yuan}
was born in Jiangsu, China in 1976. He received the B. S. and Ph.
D. degrees in electronic engineering from Zhejiang University,
Hangzhou, China, in 1999 and in 2006, respectively. He worked at
Zhejiang University from June 2006 to February 2007, and is
currently a post-doc. in Department of Electrical and Computer
Engineering, Duke University.
\end{IEEEbiographynophoto}
\begin{IEEEbiographynophoto}{Tzong-Jer Yang}
was born in Potzu, a small town of Chia-Yi Shien, Taiwan, in 1942.
He received the B.S. degree in Physics from National Taiwan Normal
University, Taipei, Taiwan, in 1964, the M.S. degree in Physics from
National Tsing Hua University, Hsinchu, Taiwan, in 1967, and Ph.D.
degree in Physics, Northwestern University, Evanston, Illinois, USA,
in 1976. He is currently Professor of Electrophysics, National Chiao
Tung University, Hsinchu, Taiwan. His research interests are in
superconductivty, first-principal calculations of metals and related
compounds, photonic crystals, waveguides in phtonic crystals,
photonic crystal fibers, near-field optics, and indefinite media.
\end{IEEEbiographynophoto}

\end{document}